\documentclass[]{jfm}
\usepackage{bm}
\usepackage{amsmath}
\usepackage{graphicx}
\newcommand{\comment}[1]{}

\ifCUPmtlplainloaded \else
  \checkfont{eurm10}
  \iffontfound
    \IfFileExists{upmath.sty}
      {\typeout{^^JFound AMS Euler Roman fonts on the system,
                   using the 'upmath' package.^^J}%
       \usepackage{upmath}}
      {\typeout{^^JFound AMS Euler Roman fonts on the system, but you
                   dont seem to have the}%
       \typeout{'upmath' package installed. JFM.cls can take advantage
                 of these fonts,^^Jif you use 'upmath' package.^^J}%
      }
  \else
  \fi
\fi

\ifCUPmtlplainloaded \else
  \checkfont{msam10}
  \iffontfound
    \IfFileExists{amssymb.sty}
      {\typeout{^^JFound AMS Symbol fonts on the system, using the
                'amssymb' package.^^J}%
       \usepackage{amssymb}%
         \let\leq=\leqslant
       \let\ge=\geqslant  
      }{}
  \fi
\fi

\ifCUPmtlplainloaded \else
  \IfFileExists{amsbsy.sty}
    {\typeout{^^JFound the 'amsbsy' package on the system, using it.^^J}%
     \usepackage{amsbsy}}
    {}
\fi

\newsavebox{\astrutbox}
\sbox{\astrutbox}{\rule[-5pt]{0pt}{20pt}}

\title[]{On the strength of the nonlinearity in isotropic turbulence}

\author[W. J. T. Bos and R. Rubinstein]%
{W\ns J.\ns  T.\ns B\ls O\ls S$^1$\ns
\and 
R.\ns R\ls U\ls B\ls I\ls N\ls S\ls T\ls E\ls I\ls N$^2$}

  \affiliation{$^1$LMFA-CNRS, Universit\'e de Lyon, Ecole Centrale de Lyon,
  69134 Ecully, France\\[\affilskip]
               $^2$Newport News, VA, USA}

\pubyear{2013}
\volume{?}
\pagerange{1--10}
\date{february 2013}
\begin{document}
\maketitle

\begin{abstract}
Turbulence governed by the Navier-Stokes equations shows a tendency to evolve towards a state in which the nonlinearity is diminished. In fully developed turbulence this tendency can be measured by comparing the variance of the nonlinear term to the variance of the same quantity measured in a Gaussian field with the same energy distribution.  In order to study this phenomenon at high Reynolds numbers, a version of the Direct Interaction Approximation  is used to obtain a closed expression for the statistical average of the mean-square nonlinearity.  The wavenumber spectrum of the mean-square nonlinear term is evaluated and its scaling in the inertial range is investigated as a function of the Reynolds number. Its scaling is dominated by the sweeping by the energetic scales, but this sweeping is weaker than predicted by a random sweeping estimate. At inertial range scales, the depletion of nonlinearity as a function of the wavenumber is observed to be constant. At large it is observed that  the mean-square 
nonlinearity is larger than its Gaussian estimate, which is shown to be related to 
the non-Gaussianity of the Reynolds-stress fluctuations at these scales.
\end{abstract}

%\pacs{47.27.Ak, 47.27.Eq, 47.27.Gs, 47.27.Jv, 47.27.Te, 47.27.Qb}

\section{Introduction}

\cite{Kraichnan1988}, observed that in turbulent flows the variance of the nonlinear term in the Navier-Stokes equations is on average smaller than would be expected from a Gaussian estimate. More precisely, if one constructs a flowfield consisting of random statistically independent Fourier modes exhibiting the same energy spectrum as the turbulent flow considered, the variance of the nonlinear term will be larger than for the original field. This depletion of nonlinearity is the result of a self-organization process of the turbulent flow, a process which is, itself, due to the nonlinear term in the Navier-Stokes equations. Kraichnan and Panda's study was motivated by the possible importance of velocity-vorticity alignment in turbulent flows, which they showed to be one expression of a more general, underlying property of nonlinear systems. We consider that this general property, the depletion of nonlinearity, is an important feature of turbulent flows, since the nonlinearity of the Navier-Stokes equations 
is the heart of the turbulence problem. 

The nonlinear term of the Navier-Stokes equations is a vector and its mean value is zero in  isotropic turbulence. An obvious question to ask is then: {\it how strong are the fluctuations of the nonlinear term and how strong is its depletion?} These questions will be addressed in the present investigation. Furthermore, we will address the following questions with respect to this phenomenon: {\it how does the depletion of nonlinearity vary as a function of the Reynolds number? Does this depletion of nonlinearity manifest itself in the inertial range? What are the physical consequences of this depletion?} 

In order to answer these questions we focus on the nonlinearity spectrum, which we will define below. This spectrum measures the strength of the fluctuations of the nonlinear term as a function of scale, just like the energy spectrum does for the strength of velocity fluctuations. Whereas the characterization of the energy spectrum has received an enormous amount of attention in the field of turbulence research, only very few investigations consider the nonlinearity spectrum. To our knowledge, only the work by \cite{Chen1989}, \cite{Nelkin1990}, and \cite{Ishihara2003} considered this quantity. \cite{Chen1989}, performed low resolution Direct Numerical Simulations and compared their results to the Direct Interaction Approximation (DIA). No information was obtained on the inertial range behavior of this quantity since the Reynolds number was too low in their simulations. Higher Reynolds numbers could in principle be obtained by using the DIA, but physically incorrect behaviour is observed in the inertial 
range dynamics of the 
original, Eulerian DIA (\cite{Kraichnan1964}). \cite{Nelkin1990}, considered only the scaling of the potential part of the advection term, assuming that the full nonlinear term scales as its potential part. Only the high resolution simulations by \cite{Ishihara2003} give an idea on the inertial range scaling of several fourth order spectra.

In the present work we use a version of the DIA in which the time-correlations are modified in order to yield results which are in agreement with Kolmogorov's inertial range phenomenology. This approach allows to investigate the strength of the nonlinearity  and its scaling properties in high Reynolds number turbulence.

\section{Inertial range scaling of the nonlinearity spectrum}

We consider the case of a unit density, incompressible, isotropic, fully developed turbulent flow governed by
\begin{eqnarray}
\frac{\partial \bm u(\bm x,t)}{\partial t}+ \bm u(\bm x,t)\cdot \nabla \bm u(\bm x,t) =-\nabla p(\bm x,t)+\nu\Delta \bm u(\bm x,t) +\bm f(\bm x,t)\\
\nabla\cdot \bm u(\bm x,t)=0 \label{eq:divu}
\end{eqnarray}
with $\bm u$ the velocity, $p$ the pressure, $\nu$ the viscosity and $\bm f$ an isotropic forcing term, confined to the largest scales of the flow. The quantities $\bm u, p$ and $\bm f$ are a function of space $\bm x$ and time $t$. The time dependence will be omitted in the following, except for quantities which depend on two or more time-instants. The Fourier transform of the velocity will be indicated by $\bm u(\bm k)$, and its evolution is given by
\begin{eqnarray}
\left[\frac{\partial}{\partial t}+\nu k^2\right] u_i(\bm k) = N_i(\bm k)+f_i(\bm k),\nonumber\\
N_i(\bm k)=-\frac{i}{2}P_{ijm}(\bm k)\iint \delta(\bm k-\bm p-\bm q)u_j(\bm p)u_m(\bm q)d\bm pd\bm q
 \end{eqnarray}
in which the pressure is eliminated using relation (\ref{eq:divu}) and where $P_{ijm}(\bm k)$ is given by
\begin{equation}
P_{ijm}(\bm k)=k_j\delta_{im}+k_m\delta_{ij}-2\frac{k_ik_jk_m}{k^2}.
\end{equation}
The energy spectrum is 
\begin{equation}
E(k)= 2\pi k^2 \left<u_i(\bm k)u_i(-\bm k)\right>,
\end{equation}
where the brackets indicate an ensemble average. The spectrum is defined such that
\begin{equation}\label{eq:defineU}
 \int E(k) dk = \frac{1}{2}\left<|\bm u(\bm x)|^2\right>\equiv \frac{1}{2}U^2.
\end{equation}
with $k=|\bm k|$. We define the wavenumber spectrum of the mean-square nonlinearity 
\begin{equation}\label{eq:NN}
w(k)=4\pi k^2\left<N_i(\bm k)N_i(-\bm k)\right>,
\end{equation}
so that
\begin{equation}
 \int w(k) dk = \left<|\bm u(\bm x)\cdot \nabla \bm u(\bm x) +\nabla p(\bm x)|^2\right> \equiv N^2.
\end{equation}
In the following $w(k)$ will be called the nonlinearity spectrum. The spectra $E(k)$ and $w(k)$ represent the distribution over scales of the kinetic energy and the mean-square nonlinearity, respectively. Omitting possible internal intermittency corrections, the kinetic energy spectrum in statistically stationary high Reynolds number isotropic turbulence scales at large $k$ as (\cite{Kolmogorov}, in the following named K41),
\begin{equation}\label{eq:K41}
 E(k)=\epsilon^{2/3}k^{-5/3}f(k\eta)
\end{equation}
with $\epsilon$ the mean dissipation rate, $\eta=\nu^{3/4}\epsilon^{-1/4}$. Here and throughout  $f(x)$ denotes a dimensionless function independent of the Reynolds number, not necessarily the same whenever it appears. If one assumes that in the inertial range the nonlinearity spectrum is likewise determined by the dissipation rate and the wavenumber, one would obtain the scaling
\begin{equation}
 w(k)=\epsilon^{4/3}k^{-1/3}f(k\eta)
\end{equation}
This is shown not to be the case and it will be shown that the scaling of $w(k)$ is more closely given by
\begin{equation}\label{eq:ScalingW}
 w(k)=U^2\epsilon^{2/3}k^{1/3}f(k\eta).
\end{equation}
where the large-scale velocity $U$ is defined in expression (\ref{eq:defineU}). This scaling implies, by integration of the latter expression up to $k_\eta\equiv 2\pi/\eta$, that the mean-square nonlinearity varies as 
\begin{equation}\label{eq:N2}
 N^2\sim R_\lambda^2,
\end{equation}
for asymptotically high Reynolds numbers ($R_\lambda$ is the Taylor-scale Reynolds number). It will furthermore be shown that $N^2/(N^2)^G$ tends to a non-unity value, independent of the Reynolds number. This implies that not only $w(k)$, but also it Gaussian estimate scales as
\begin{equation}\label{eq:ScalingWG}
 w^G(k)=U^2\epsilon^{2/3}k^{1/3}f(k\eta),
\end{equation}
where $w^G(k)$ is the nonlinearity spectrum computed from the same velocity field, assuming independence of the Fourier modes.

In the following we will try to establish the scaling expressions (\ref{eq:ScalingW}), (\ref{eq:N2}) and (\ref{eq:ScalingWG}), and we will show how non-Gaussian effects influence this scaling. In order to show clear scaling ranges,  high Reynolds numbers are needed.  We derive a DIA expression for the nonlinearity spectrum, which we simplify to obtain an expression of the Eddy-Damped Quasi-Normal Markovian (EDQNM) type. This derivation is presented in section \ref{sec:Closure}. The resulting expression for the nonlinearity spectrum and its Gaussian estimate are functionals of the energy spectrum, wavenumber and viscosity only,
\begin{eqnarray}
 w(k)=F[E(k),k,\nu]\nonumber\\
 w^G(k)=F[E(k),k],
\end{eqnarray}
that is, once we prescribe the energy spectrum and the Reynolds number, we can evaluate $w(k)$ and $w^G(k)$. In section \ref{sec:Results} we will perform a numerical integration of the EDQNM closure of the Lin-equation for the energy spectrum. The hereby obtained energy spectrum is then  used to evaluate $w(k)$, $w^G(k)$, $\int w(k) dk$ and $\int w^G(k) dk$, and the dependence of these quantities on the wavenumber and Reynolds number is investigated. In section \ref{sec:SuperGaussian} the large-scale behaviour of the nonlinearity spectrum and its link with the Reynolds-stress fluctuation spectrum is discussed.  

\section{Gaussian estimate and closure expression for the nonlinearity spectrum \label{sec:Closure}}

\subsection{Gaussian estimate of the mean-square nonlinearity: random sweeping \label{sec:Gauss}}

Evaluating $w(k)$, as defined by (\ref{eq:NN}) assuming independence of the Fourier modes yields the Gaussian estimate ({\it cf.} \cite{Chen1989}),
\begin{equation}
  w^G(k)=k^3 \iint_\Delta a(k,p,q) E(p)E(q)\frac{dp~dq}{pq}\label{eq:wgk}.
\end{equation}
The symbol $\Delta$ indicates the domain in the $pq$-plane in which
$k,p,q$ can form a triangle (in other words $|p-q|\leq k \leq|p+q|$), the quantity  $a(k,p,q)$ is given by
\begin{equation}
a(k,p,q)=\frac{1}{2}(1-xyz-2y^2z^2)
\end{equation}
and $x,y,z$ are 
\begin{eqnarray}
x=-p_iq_i/(pq)\nonumber\\
y=k_iq_i/(kq)\nonumber\\
z=k_ip_i/(kp).\label{eq:xyz}
\end{eqnarray}
The Gaussian estimate of the nonlinearity spectrum is thus completely determined once the energy spectrum is given.  
Considering in some detail expression (\ref{eq:wgk}), and in particular the quantity $a(k,p,q)$, it is observed that the integral is dominated by triad interactions in which $k\approx p\gg q$ and $k\approx q\gg p$. For instance, when $k\approx q\gg p$, $x\approx z\approx 0$ and $y\approx 1$, so that $a(k,p,q)$ is not zero, and contributions from the infrared range will determine the integral. This allows to obtain the following approximation,
\begin{eqnarray}\label{eq:ScalingGauss0}
w^G(k)&\sim& k^2 E(k) \int E(p)dp 
\end{eqnarray}
which, assuming K41 scaling, yields
\begin{equation}\label{eq:ScalingGauss}
w^G(k)\sim U^2 \epsilon^{2/3} k^{1/3}.
\end{equation}
We can thus analytically establish the scaling for $w^G(k)$. Note that a simple dimensional analysis, based on the observation that the nonlinear transfer is dominated by sweeping, and proportional to the spectrum of the Eulerian velocity gradient, $k^2 E(k)$ gives the same expression (\ref{eq:ScalingGauss}). 
 This  analysis (which is a formulation of Tennekes' random sweeping estimate (\cite{Tennekes1975})) implicitly assumes independence of the Fourier modes at different scales and is thus equivalent to the Gaussian estimate. In a true turbulent field in which the modes are not independent, this analysis is not {\it a priori} satisfied, and how the dependence between Fourier modes will alter this scaling, {\it i.e.}, how the cumulant contributions to $w(k)$ scale, will be considered in the following section.

\subsection{Direct interaction approximation for the mean-square nonlinearity \label{sec:DIA}}

The Direct Interaction Approximation (\cite{KraichnanDIA}) allows to investigate the influence of the inter-dependence of the Fourier modes in a turbulent flow under the assumption that the individual coupling between the triads is weak. The collective influence of the coupling of all triads together is however not necessarily weak and the DIA can consider systems which are far from Gaussianity. The obtained results by the original, Eulerian, DIA are not invariant under random Galilean transformations, which is not in agreement with the physics of a turbulent flow. This manifests itself by the fact that the Eulerian DIA yields an energy spectrum which is not in agreement with Kolmogov's scaling phenomenology, equation (\ref{eq:K41}). In order to cure for this, the DIA can be formulated in Lagrangian coordinates (\cite{Kraichnan1964}). The resulting set of equations (\cite{Kraichnan65}) is complicated (see \cite{Kaneda81}, for a more tractable variant of Lagrangian DIA) and depends on the entire history of 
the flow. Our approach to analyze the effects of dependence of the modes avoids these problems, since we start from Eulerian DIA and explicitly model the time-dependence of the Fourier modes (\cite{KraichnanTFM}). The correlation time that we will use in the time-correlation functions is chosen such that the results are consistent with Kolmogorov's scaling arguments.

A straightforward way to derive the closure expression for the mean-square nonlinearity is by using the generalized Langevin model for the Direct Interaction Approximation (\cite{Kraichnan1970,Leith2}). This approach was also described in \cite{Chen1989}, and we used this approach to derive a closed expression for the mean-square advection term in \cite{Bos2012-3}. The DIA Langevin model is given by 
\begin{eqnarray}\label{eq:Langevin}
 \left[\frac{\partial}{\partial t}+\nu k^2\right] u_i(\bm k,t)&=&q_i(\bm k,t)-\int_0^t\eta(k,t,s)u_i(\bm k,s)ds\\
&=&q_i(\bm k,t)-d_i(\bm k,t)
\end{eqnarray}
with
\begin{eqnarray}
q_i(\bm k,t)=-\frac{i}{2}P_{ijm}(\bm k)\int \delta(\bm k-\bm p-\bm q) \zeta_j(\bm p,t)\zeta_m(\bm q,t)d\bm pd\bm q\\
d_i=\int_0^t\eta(k,t,s)u_i(\bm k,s)ds\\
\eta(k,t,s)=\frac{1}{2}\int_\Delta  k p^2 b(k,p,q)G(p,t,s)E(q,t,s)\frac{dp}{p}\frac{dq}{q}.
\end{eqnarray}
where $\zeta_i(\bm k,t)$ is an independent Gaussian random variable with the same two-time correlation function as $u_i$, 
\begin{equation}
E(k,t,t')=2\pi k^2\left<u_i(\bm k,t)u_i(-\bm k,t')\right>=2\pi k^2\left<\zeta_i(\bm k,t)\zeta_i(-\bm k,t')\right>. 
\end{equation}
$G(k,t,s)$ is the response function (or Green's function) and $b(k,p,q)=(p/k)(xy+z^3)$.  
The term $d_i$ is the damping term of the Langevin equation, where the damping is due to nonlinear scrambling, or eddy damping, and viscous effects.
Since (\ref{eq:Langevin}) is a linear function of $u_i$, it can be inverted, giving
\begin{equation}\label{eq:Inverted}
u_i(\bm k,t)=\int_0^t G(k,t,s) q_i(\bm k,s)ds.
\end{equation}

The spectrum of the mean-square nonlinearity is given by the square of the RHS of equation \ref{eq:Langevin},
\begin{equation}\label{eq:Langevin2}
 w(k,t)=4\pi k^2 \left<\left|q_i(\bm k,t)-d_i(\bm k,t)\right|^2\right>.
\end{equation}
The different terms that appear are then,
\begin{eqnarray}\label{eq:MSNL2}
 \left<\left|q_i(\bm k,t)\right|^2\right>&=&\frac{1}{4}P_{ijm}(\bm k)P_{iln}(\bm k)\times\nonumber\\
&&\int \delta_{\bm k-\bm p-\bm q}\delta_{\bm k-\bm p'-\bm q'}\left< \zeta_j(\bm p,t)\zeta_m(\bm q,t)\zeta_l(-\bm p',t)\zeta_n(-\bm q',t)\right>d\bm p' d\bm q' d\bm pd\bm q\\
\left<\left|d_i(\bm k,t)\right|^2\right>&=&\int_0^t\int_0^t\eta(k,t,s)\eta(k,t,s')\left<u_i(\bm k,s)u_i(-\bm k,s')\right>ds'ds\\
\left<q_i(\bm k,t)d_i^*(\bm k,t)\right>&=&q_i(\bm k,t)\int_0^t\eta(k,t,s)u_i(\bm k,s)ds.~~~~~~~~~~~~~~~~~~~~~~~~~~~~~~~~~~~~~~~~~~~~~~~~~~~
\end{eqnarray}
The first term can be simplified by using the rules for Gaussian quantities, leading to
\begin{eqnarray}\label{eq:MSNL3}
 \left<\left|q_i(\bm k,t)\right|^2\right>=\frac{w^G(k)}{4\pi k^2}
\end{eqnarray}
with $w^G(k)$ given by (\ref{eq:wgk}). The second term is directly closed. The last term can be closed by using expression (\ref{eq:Inverted}), yielding
\begin{eqnarray}\label{eq:MSNL4}
\left<q_i(\bm k,t)d_i^*(\bm k,t)\right>&=&\int_0^t \int_0^s \eta(k,t,s)G(k,s,s')\left<q_i(\bm k,t)q_i(-\bm k,s')\right>ds' ds.
\end{eqnarray}
The resulting expression for $w(k)$ is then,
\begin{small}
\begin{eqnarray}\label{eq:MSNL5}
&&w(k)=w^G(k)+
\nonumber\\
&&\frac{1}{2}\int_0^t\int_0^t \int_\Delta\int_{\Delta'} bb' k^2p^2p'^2 G(p,t,s)G(p',t,s')E(q,t,s)E(q',t,s')E(k,s,s')
\frac{dp'dq'}{p'q'}\frac{dp~dq}{pq}ds'ds\nonumber\\
&&-\int_0^t\int_0^s \int_\Delta\int_{\Delta'} bb' k^4p^2 G(p,t,s')G(k,s,s')E(q,t,s)E(q',t,s')E(p',t,s')
\frac{dp'dq'}{p'q'}\frac{dp~dq}{pq}ds'ds\nonumber\\
\end{eqnarray}
\end{small}
with $b=b(k,p,q)$ and $b'=b(k,p',q')$. This expression, which is a function of the response function and the energy spectrum only, is closed within the DIA formalism. 

\subsection{Single-time expressions}\label{secMarkov}

As was discussed in the beginning of section \ref{sec:DIA}, DIA is a two-time theory which is not compatible with Kolmogorov phenomenology. In order to obtain a single-time model which is, we introduce simplifying assumptions, which are now discussed. 

First, to simplify expression (\ref{eq:MSNL5}) we assume an exponential Lagrangian decorrelation of the Fourier modes, with a wavenumber-dependent characteristic correlation time which needs to be defined. We will use the following expressions
\begin{eqnarray}
 G(k,t,s)=e^{-\eta(k)(t-s)}H(t-s)\nonumber\\
 E(k,t,s)=E(k)\left[G(k,t,s)+G(k,s,t)\right],
\end{eqnarray}
which assumes that the fluctuation-dissipation theorem holds for the Lagrangian velocity fluctuations. Substituting these expressions in (\ref{eq:MSNL5}), one obtains, using $(p,q)\leftrightarrow (p',q')$ symmetry
\begin{eqnarray}\label{eq:MSNL6}
w(k)=w^G(k)+\int_\Delta\int_{\Delta'} bb' k^4p^2p'^2 E(q)E(q',t)\Xi(k,p,q,p',q')
\times\nonumber\\
\left[\frac{E(k)}{k^2}-\frac{E(p')}{p'^2}\right]
\frac{dp'dq'}{p'q'}\frac{dp~dq}{pq},
\end{eqnarray}
with 
\begin{eqnarray}
\Xi(k,p,q,p',q')=\int_0^t\int_0^s G(p,t,s)G(p',t,s')G(q,t,s)G(q',t,s')G(k,s,s').
\end{eqnarray}
We have written (\ref{eq:MSNL6}) in a form from which it is directly observed that the cumulant part vanishes in thermal equilibrium, when $E(k)\sim k^2$, independent from the form of $\Xi(k,p,q,p',q')$, as was also shown for the passive scalar case in \cite{Bos2012-3}. Some qualitative predictions can also be obtained from this expression. If the integral in the inertial range is dominated by interactions in which $p'<k$, the cumulant contribution will be negative. Logically, for the smallest wavenumbers $k$ the integral can not be dominated by  $p'<k$, so that interactions with $p'\ge k$ must determine the integral. For these interactions the cumulant contribution is positive. It will be seen in the following that these qualitative predictions are indeed in agreement with the results of the numerical integration of expression (\ref{eq:MSNL6}).

Working out the integrals we find,
\begin{eqnarray}
\Xi(k,p,q,p',q')=\frac{1}{A_{pqp'q'}-B_{kp'q'}}\left[\frac{1-e^{-B_{kp'q'}t}}{B_{kp'q'}}-\frac{1-e^{-A_{pqp'q'}t}}{A_{pqp'q'}}\right]\nonumber\\
A_{pqp'q'}=\eta_p+\eta_q+\eta_{p'}+\eta_{q'}\nonumber\\
B_{kp'q'}=\eta_k+\eta_{p'}+\eta_{q'}.
\end{eqnarray}
Note that for long times this expression simplifies to
\begin{eqnarray}
\Xi(k,p,q,p',q')=\frac{1}{A_{pqp'q'}B_{kp'q'}}.
\end{eqnarray}
In this expression we need to define the  eddy damping frequency $\eta_k^{-1}$. We will use a response frequency which is compatible with a $k^{-5/3}$ inertial range and with a dominant viscous contribution at large wavenumbers,
\begin{eqnarray}
\eta_k=\lambda \sqrt{\int_0^k s^2 E(s)ds} +\nu k^2.
\end{eqnarray}
The constant $\lambda$ is chosen $0.5$. In a long K41 inertial range $\eta_k$ is proportional to $\epsilon^{1/3}k^{2/3}$. This choice for the eddy-damping is common in the EDQNM model, where the damping can be freely chosen. A more sophisticated closure is the Test Field Model (\cite{KraichnanTFM}) or the LMFA closure (\cite{Bos2013-1}), where the damping is determined self-consistently by solving an additional equation for an advected test-field. It is expected that the use of these closures to determine the damping will not qualitatively change the results, since the inertial range behavior of $\eta_k$ is qualitatively similar.

\section{Results for the mean-square nonlinearity \label{sec:Results}}

\subsection{EDQNM results for the energy spectrum}

\begin{figure}\begin{center}
\includegraphics[width=0.5\linewidth]{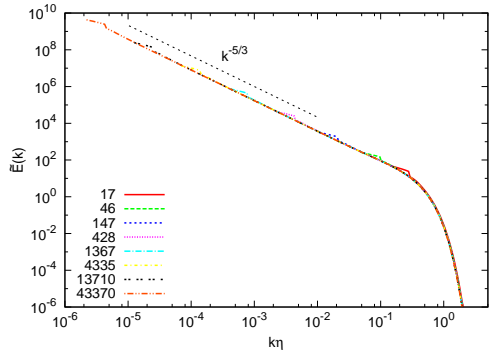}~
\caption{Energy spectrum, normalized by Kolmogorov variables $\tilde{E}(k)=E(k)/(\epsilon^{1/4}\nu^{5/4})$. 
\label{Fig:Ek}}
\end{center}\end{figure}

 We use here an energy spectrum obtained using the EDQNM model (\cite{Orszag}), and details on the method and discretization can be found in \cite{Bos2012-3}. The velocity field is forced at the largest scales, and the energy spectrum is evaluated once a steady state is obtained. The energy spectra for $17<R_\lambda <4\cdot 10^4$ are shown in Figure \ref{Fig:Ek}. Using Kolmogorov variables (length- and timescales constructed using the dissipation rate $\epsilon$ and viscosity $\nu$), the spectra collapse perfectly in the inertial and dissipation range. We can only distinguish the different spectra in their forcing range. 

\subsection{Reynolds number dependence and Gaussianity of the mean-square nonlinearity}

\begin{figure}
\includegraphics[width=0.5\linewidth]{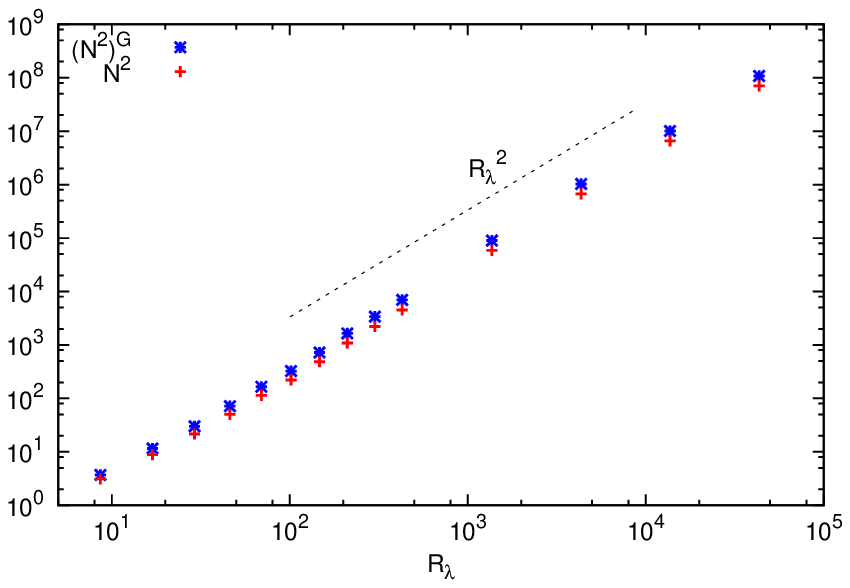}~
\includegraphics[width=0.5\linewidth]{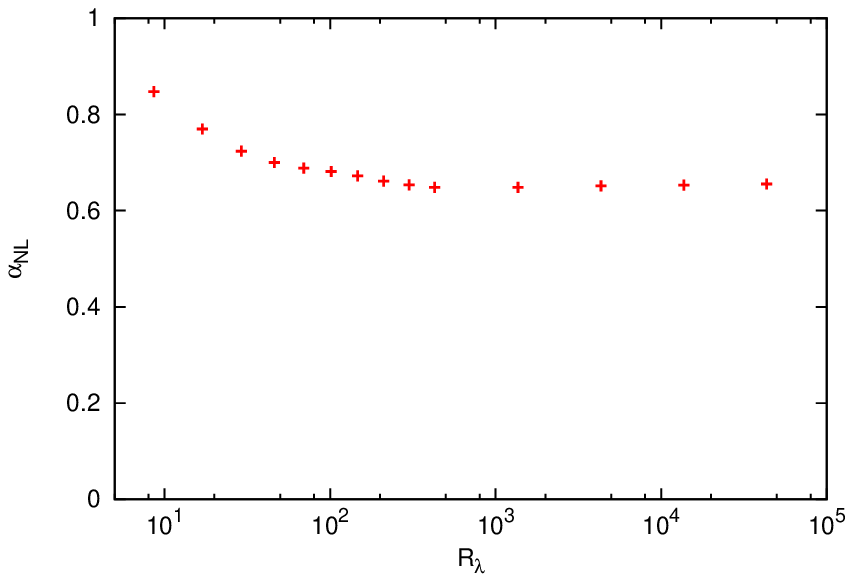}
\caption{Left: mean-square nonlinearity $N^2$ and its Gaussian estimate $(N^2)^G$ as a function of the Reynolds number. Right: depletion of nonlinearity, quantified by $\alpha_{NL}=N^2/(N^2)^G$, as a function of the Reynolds number. \label{Fig:Wint}}
\end{figure}

In Figure \ref{Fig:Wint}, the mean-square nonlinearity and its Gaussian estimate, computed from expressions (\ref{eq:MSNL6}) and (\ref{eq:wgk}) are shown as a function of the Reynolds number. Both $N^2$ and $(N^2)^G$ increase proportional to $R_\lambda^2$. In Figure \ref{Fig:Wint}, right, it is shown that the ratio between the two quantities tends to a constant value, which is approximately $0.65$ for large $R_\lambda$, a value which is rapidly approached for $R_\lambda>100$. This value is of the same order as observed in simulations of low Reynolds number decaying turbulence (\cite{Kraichnan1988}).

\subsection{Scaling of the nonlinearity spectrum}

\begin{figure}
\includegraphics[width=0.5\linewidth]{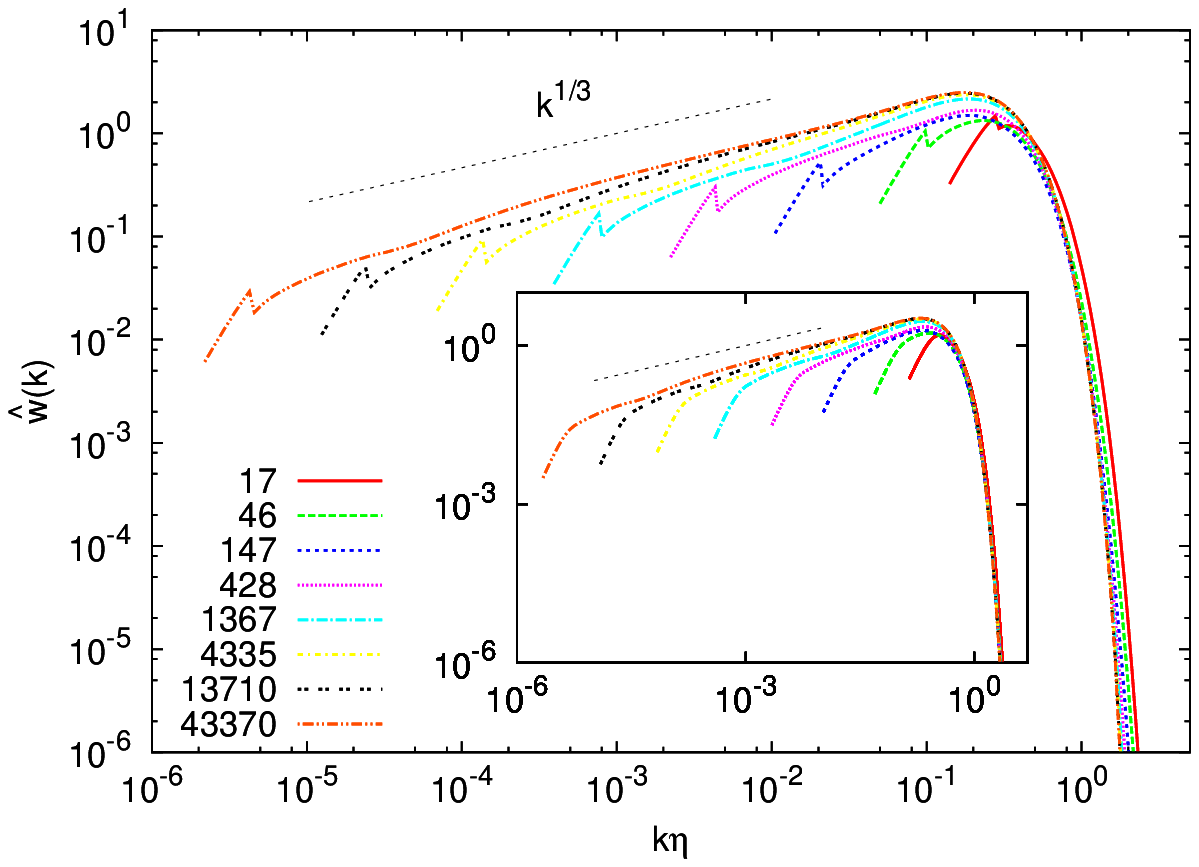}~
\includegraphics[width=0.5\linewidth]{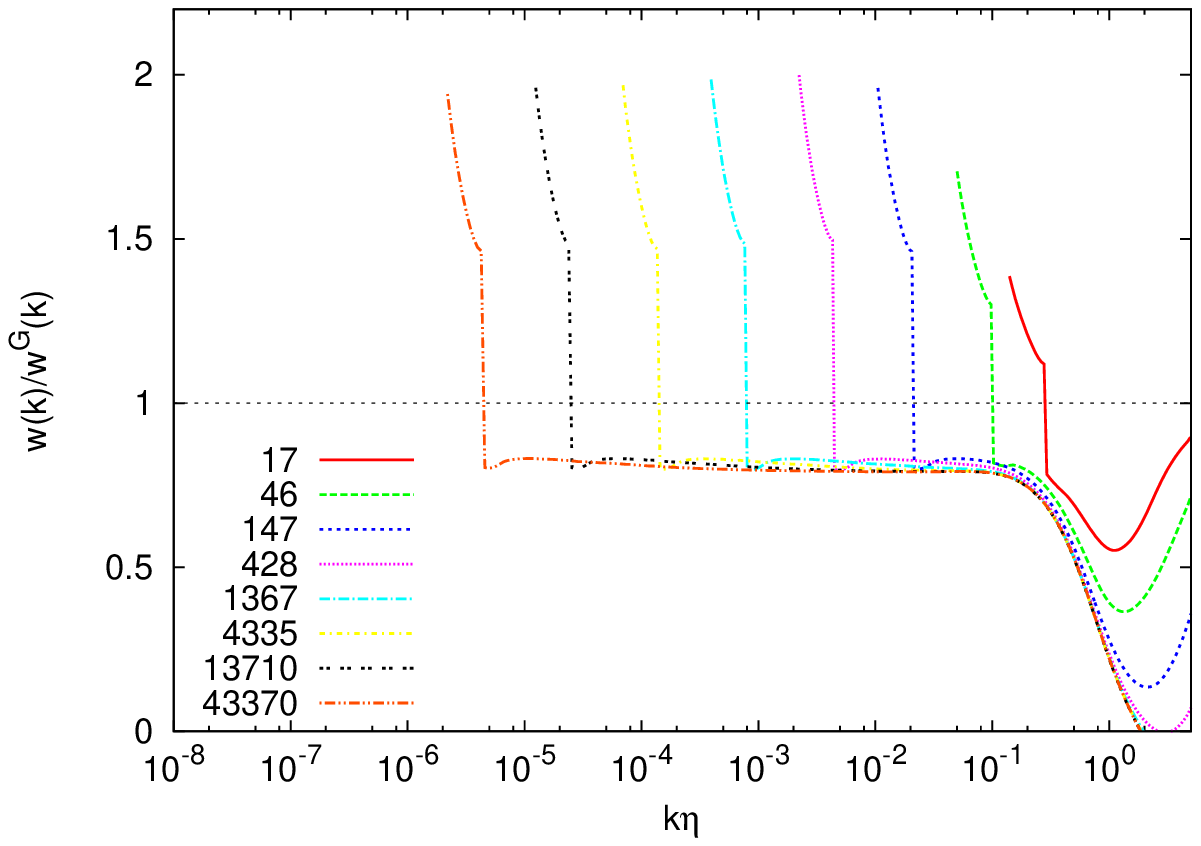}
\caption{Left: nonlinearity spectrum normalized by sweeping variables, $\tilde{w}(k)=w(k)/(U^2\epsilon^{3/4}\nu^{-1/4})$. Inset: normalized Gaussian nonlinearity spectrum $\tilde{w}^G(k)$.  Right: spectrum of the nonlinear term divided by its Gaussian estimate.\label{Fig:Wk}}
\end{figure}

In Figure \ref{Fig:Wk} (left), we show the nonlinearity spectra normalized by 'sweeping-variables'
\begin{eqnarray}
 \tilde{w}(k)=w(k)/(U^2\epsilon^{3/4}\nu^{-1/4}).
\end{eqnarray}
We observe that the spectra collapse at the high wavenumbers, even though the superposition is not as perfect as for the energy spectra (Figure \ref{Fig:Ek}). However, to a good approximation, the spectra scale as
\begin{eqnarray}
 w(k)\sim U^2\epsilon^{2/3}k^{1/3}f(k\eta)
\end{eqnarray}
with $f(k\eta)$ a function which tends to a constant in the inertial range and which rapidly decays in the dissipation range. Our simulations do thus confirm that both $w(k)$ and $w^G(k)$ scale proportional to $U^2\epsilon^{2/3}k^{1/3}$ in the inertial range. A clear power-law scaling proportional to $k^{1/3}$ appears, however, only at relatively high Reynolds number. For moderate and low Reynolds numbers the power law is steeper, {\it i.e.} the power-law exponent is larger than $1/3$.  The inherent nonlocal character of the sweeping contribution to the nonlinear term might be behind the slow convergence to an asymptotic inertial range scaling.

In Figure \ref{Fig:Wk} (right), we plot the ratio of the nonlinearity spectrum to its Gaussian estimate, $w(k)/w^G(k)$. This representation shows directly how the nonlinearity spectrum is affected at different scales by the cumulant contributions. It is observed that the nonlinearity spectrum is super-Gaussian in the forced scales. Here and in the following we mean by sub- or super-Gaussian that the value of a quantity in the turbulent flow is smaller or larger, respectively, than its value in the Gaussian reference field. In the inertial range a constant depletion of nonlinearity is observed. This implies that at these scales the cumulant spectrum scales exactly like the Gaussian spectrum as a function of wavenumber. In the dissipation range the cumulant contribution becomes more strongly negative, leading to a more important depletion of nonlinearity in these scales. In the far dissipation range the spectrum of the nonlinear term shows negative values for high Reynolds numbers. This non-realizable 
behavior for $k\eta>2$ might be due to the numerical integration of the expressions, or to the procedure we used to obtain single-time expressions, which does not guarantee the preservation of the realizability property of the Eulerian DIA.

At the inertial range scales the depletion of nonlinearity is thus approximately constant and the statistics for the mean-square nonlinearity are sub-Gaussian. The fact that the mean-square nonlinearity is weaker than its Gaussian estimate indicates that the effect of random sweeping is reduced, and suggests a certain order at these scales. How the amount of order is determined, \emph{i.e.}, what determines the level of depletion of nonlinearity, is currently under investigation. The super-Gaussian behaviour in the large scales will be addressed in the following section.

\section{Fluctuations of the Reynolds stress and super-Gaussian behaviour of the large-scale nonlinearity fluctuations\label{sec:SuperGaussian}}

Using DIA techniques, expressions can be derived for the cumulant contributions to all sorts of correlations at arbitrary order. Some details on the procedure are given in \cite{Chen1989}. It was shown that, whereas the cumulant contributions to the mean-square nonlinearity spectrum are nonzero within the DIA framework, fourth-order vorticity correlations, dissipation rate fluctuations and pressure-gradient fluctuations are all Gaussian according to the DIA, which seems to be in disagreement with observations from direct numerical simulations. The precise reason for this is still not clear at present. In addition to the expression for the mean-square nonlinearity, we derived expressions for the cumulants to the correlations

\begin{small}
\begin{equation}
T_{ijij}(k)=4\pi k^2\int  \delta(\bm k-\bm p-\bm q)\delta(\bm k-\bm p'-\bm q')\left<u_i(\bm p)u_j(\bm q)u_i(-\bm p')u_j(-\bm q')\right>d\bm pd\bm qd\bm p'd\bm q'
\end{equation}
\end{small}
and

\begin{small}
\begin{equation}
T_{iijj}(k)=4\pi k^2\int \delta(\bm k-\bm p-\bm q)\delta(\bm k-\bm p'-\bm q')\left<u_i(\bm p)u_i(\bm q)u_j(-\bm p')u_j(-\bm q')\right>d\bm pd\bm qd\bm p'd\bm q'.
\end{equation}
\end{small}
The first of these two expressions is the Reynolds stress fluctuation spectrum, which gives a measure for the fluctuations of the Reynolds stress at various scales. The second,  $T_{iijj}(k)$ is the energy-fluctuation spectrum, which measures the fluctuations of the kinetic energy at various scales. After some long but straightforward algebra, one obtains that
\begin{eqnarray}
T_{iijj}^C(k)=0,
\label{eq:TijijW}\nonumber\\
T_{ijij}^C(k)=\frac{2}{k^{2}} w^C(k),
\end{eqnarray}
where the superscript $C$ indicates that we consider the cumulant part of the spectrum. The cumulant part of the nonlinearity spectrum $w^C(k)$ is given by the second and third line of expression (\ref{eq:MSNL5}). This shows that according to DIA the energy fluctuation spectrum is given by its Gaussian estimate, whereas the Reynolds-stress fluctuation spectrum is not. It also shows that the non-Gaussian part of $T_{ijij}(k)$ is related, in a simple way, to the nonlinearity spectrum.

Since, evidently, the single point correlation $\left<u_iu_iu_ju_j\right>=\left<u_iu_ju_iu_j\right>$, we must have
\begin{equation}
\int T_{ijij}(k) dk =\int T_{iijj}(k) dk
\end{equation}
and therefore
\begin{equation}
\label{eq:Tijij0}
\int T_{ijij}^C(k) dk =\int T_{iijj}^C(k)dk=0.
\end{equation}
The spectrum of the cumulant to the Reynolds-stress fluctuations $T_{ijij}^C(k)$  contains thus, if it is non-zero, both positive and negative contributions, which sum up to zero when integrated over all scales. This is comparable to the nonlinear transfer spectrum, which also sums up to zero. According to relation (\ref{eq:TijijW}) we can link the spectrum $T_{ijij}^C(k)$ to the nonlinearity spectrum. We have seen (for example in Figure \ref{Fig:Wk}) that a negative cumulant-contribution is observed in the inertial and dissipation range scales for $w(k)$. This holds thus also for $k^{-2}w^C(k)$. This negative contribution must be compensated by a positive contribution in the large scales in order to satisfy (\ref{eq:Tijij0}). The super-Gaussian statistics in the nonlinearity spectrum at large scales are therefore, within the DIA approach, directly linked to the depletion of nonlinearity in the small scales, through the relations (\ref{eq:TijijW}) and (\ref{eq:Tijij0}), involving the Reynolds stress 
fluctuation spectrum.

\section{Discussion and conclusion}

In the present investigation we have considered and established the wavenumber scaling and Reynolds number scaling of the mean-square nonlinearity. It is shown that, in the inertial and dissipation range, the nonlinearity spectrum is given by
\begin{eqnarray}\label{eq:wconc}
 w(k)= U^2\epsilon^{2/3}k^{1/3}f(k\eta),
\end{eqnarray}
for very high Reynolds numbers. The function $f(k\eta)$ tends to a constant value in the inertial range and its value is approximately $0.8$ times the value of its Gaussian estimate. The total depletion of nonlinearity, measured by the ratio of $N^2$ to $(N^2)^G$ is shown to tend to a constant value of approximately $0.65$. This sub-Gaussian behavior of turbulence must be connected with a certain order in the flows, but how this manifests itself in an instantaneous flow field ({\it e.g.} in terms of coherent flow structures) cannot be inferred from the statistical considerations presented here. 

The nonlinear term consists of two parts: the advection term and the pressure gradient term. Since the pressure spectrum, $E_p(k)$ scales approximately as $E_p(k)\sim \epsilon^{3/4}k^{-7/3}f(k\eta)$ (\cite{Gotoh2001}), the pressure gradient spectrum scales as
\begin{eqnarray}\label{eq:Ep}
 E_{\nabla p}(k)\sim \epsilon^{4/3}k^{-1/3}f(k\eta).
\end{eqnarray}
Note however that this scaling appears only at relatively high Reynolds number (\cite{Gotoh2001}) compared to the appearance of K41 scaling for the energy spectrum.
Considering equation (\ref{eq:wconc}) and (\ref{eq:Ep}) it is clear that at large Reynolds numbers $N^2$ is only weakly determined by the variance of the pressure gradient. The variance of the nonlinearity is therefore dominantly determined by the advection term. The depletion of nonlinearity implies hereby directly a depletion of the sweeping compared to the kinematic sweeping induced by a field consisting of independent Fourier modes.
In this context we refer to the work by \cite{Chen1989-2}, which discusses the possibility of a reduction of sweeping in turbulence. They argue that a complete reduction of sweeping is improbable for stochastically forced Navier-Stokes turbulence. Their arguments are not in disagreement with the present investigation.
 The dependence of the large and small scales is  influenced, and the sweeping, as estimated by purely kinematic arguments, is partially but definitely not completely suppressed. In this light the depletion of nonlinearity can also be interpreted as a reduction of Eulerian acceleration, suggesting a larger Eulerian coherence for turbulence than for advection by random Fourier modes. The possible link of this enhanced coherence with inertial range and dissipation range intermittency is not clear at present. The super-Gaussian values of the large-scales of the nonlinearity spectrum were shown to be related to the non-Gaussianity of the Reynolds-stress-fluctuation spectrum. The physical importance of this relation for the dynamics of turbulent flow seems to deserve further research.

We mention here that a similar picture (large-scale super-Gaussian behaviour and sub-Gaussian inertial range and dissipation range behavior), was observed in the depletion of advection (\cite{Bos2012-3}), where the inertial range scaling of the advection spectrum also displayed a constant reduction with respect to its Gaussian value. An interesting perspective is the analysis of the scale distribution of the nonlinearity in magnetohydrodynamics, a system in which it was recently shown that the nonlinearity is also depleted (\cite{Servidio2008}).

%\bibliographystyle{report}
%\bibliographystyle{myJFM}
%\bibliography{/home/bos/PUBLI/biblio}

\begin{thebibliography}{}

\bibitem[Bos \& Bertoglio, 2013]{Bos2013-1}
\sc Bos, W. \rm \& \sc Bertoglio, J.-P. \rm 2013.
\newblock Lagrangian Markovianized Field Approximation for turbulence.
\newblock {\em J. Turbul.}, ~14, 99.

\bibitem[Bos {\it et~al.}, 2012]{Bos2012-3}
\sc Bos, W.J.T. \rm, \sc Rubinstein, R. \rm, \& \sc Fang, L. \rm 2012.
\newblock Reduction of mean-square advection in turbulent passive scalar
  mixing.
\newblock {\em Phys. Fluids}, ~24, 075104.

\bibitem[Chen {\it et~al.}, 1989]{Chen1989}
\sc Chen, H. \rm, \sc Herring, J.R. \rm, \sc Kerr, R.M. \rm, \& \sc Kraichnan,
  R.H. \rm 1989.
\newblock Non-Gaussian statistics in isotropic turbulence.
\newblock {\em Phys. Fluids A}, ~1, 1844.

\bibitem[Chen \& Kraichnan, 1989]{Chen1989-2}
\sc Chen, S. \rm \& \sc Kraichnan, R.H. \rm 1989.
\newblock Sweeping decorrelation in isotropic turbulence.
\newblock {\em Phys. Fluids A}, ~1, 2019.

\bibitem[Gotoh \& Fukayama, 2001]{Gotoh2001}
\sc Gotoh, T. \rm \& \sc Fukayama, D. \rm 2001.
\newblock Pressure Spectrum in Homogeneous Turbulence.
\newblock {\em Phys. Rev. Lett.}, ~86, 3775.

\bibitem[Ishihara {\it et~al.}, 2003]{Ishihara2003}
\sc Ishihara, T. \rm, \sc Kaneda, Y. \rm, \sc Yokokawa, M. \rm, \sc Itakura, K.
  \rm, \& \sc Uno, A. \rm 2003.
\newblock Spectra of Energy Dissipation, Enstrophy and Pressure by
  High-Resolution Direct Numerical Simulations of Turbulence in a Periodic Box.
\newblock {\em J. Phys. Soc. Japan}, ~72.

\bibitem[Kaneda, 1981]{Kaneda81}
\sc Kaneda, Y. \rm 1981.
\newblock Renormalized expansions in the theory of turbulence with the use of
  the {L}agrangian position function.
\newblock {\em J. Fluid. Mech.},  107, 131 -- 145.

\bibitem[Kolmogorov, 1941]{Kolmogorov}
\sc Kolmogorov, A.N. \rm 1941.
\newblock The local structure of turbulence in incompressible viscous fluid for
  very large {R}eynolds numbers.
\newblock {\em Dokl. Akad. Nauk. SSSR}, ~30, 301.

\bibitem[Kraichnan, 1959]{KraichnanDIA}
\sc Kraichnan, R.H. \rm 1959.
\newblock The structure of isotropic turbulence at very high {R}eynolds
  numbers.
\newblock {\em J. Fluid Mech.}, ~5, 497--543.

\bibitem[Kraichnan, 1964]{Kraichnan1964}
\sc Kraichnan, R.H. \rm 1964.
\newblock Kolmogorov's Hypotheses and {E}ulerian Turbulence Theory.
\newblock {\em Phys. Fluids}, ~7, 1723.

\bibitem[Kraichnan, 1965]{Kraichnan65}
\sc Kraichnan, R.H. \rm 1965.
\newblock Lagrangian-History Closure Approximation for Turbulence.
\newblock {\em Phys. Fluids}, ~8, 575.

\bibitem[Kraichnan, 1970]{Kraichnan1970}
\sc Kraichnan, R.H. \rm 1970.
\newblock Convergents to turbulence functions.
\newblock {\em J. Fluid Mech.}, ~41, 189.

\bibitem[Kraichnan, 1971]{KraichnanTFM}
\sc Kraichnan, R.H. \rm 1971.
\newblock An almost-{M}arkovian {G}alilean-invariant turbulence model.
\newblock {\em J. Fluid Mech.}, ~47, 513.

\bibitem[Kraichnan \& Panda, 1988]{Kraichnan1988}
\sc Kraichnan, R.H. \rm \& \sc Panda, R. \rm 1988.
\newblock Depression of nonlinearity in decaying isotropic turbulence.
\newblock {\em Phys. Fluids}, ~31, 2395.

\bibitem[Leith, 1971]{Leith2}
\sc Leith, C.E. \rm 1971.
\newblock Atmospheric predictability and two-dimensional turbulence.
\newblock {\em J. Atmos. Sci.}, ~28, 145--161.

\bibitem[Nelkin \& Tabor, 1990]{Nelkin1990}
\sc Nelkin, M. \rm \& \sc Tabor, M. \rm 1990.
\newblock Time correlations and random sweeping in isotropic turbulence.
\newblock {\em Phys. Fluids A}, ~2, 81.

\bibitem[Orszag, 1970]{Orszag}
\sc Orszag, S.A. \rm 1970.
\newblock Analytical theories of Turbulence.
\newblock {\em J. Fluid Mech.}, ~41, 363.

\bibitem[Servidio {\it et~al.}, 2008]{Servidio2008}
\sc Servidio, S. \rm, \sc Matthaeus, W.~H. \rm, \& \sc Dmitruk, P. \rm 2008.
\newblock Depression of Nonlinearity in Decaying Isotropic MHD Turbulence.
\newblock {\em Phys. Rev. Lett.},  100, 095005.

\bibitem[Tennekes, 1975]{Tennekes1975}
\sc Tennekes, H. \rm 1975.
\newblock Eulerian and {L}agrangian time microscales in isotropic turbulence.
\newblock {\em J. Fluid Mech.}, ~67, 561.

\end{thebibliography}

%\newpage

\end{document}